\documentclass[%
  aps,
  preprint,
  floatfix,
  amsmath,
  amssymb
]{revtex4-2}

\usepackage{graphicx}
\usepackage{dcolumn}
\usepackage{bm}
\renewcommand*{\d}{\ensuremath{\mathrm{d}}} 
\newcommand*{\ii}{\mathrm{i}} 

\begin{document}


\title{Classical and quantum dynamics of a particle confined in a paraboloidal cavity}

\author{Ángel E. Reyna-Cruz}

\author{Julio C. Guti\'errez-Vega}
\email{juliocesar@tec.mx}
\affiliation{Photonics and Mathematical Optics Group, Tecnol\'ogico de Monterrey, Monterrey 64849, M\'exico}

\begin{abstract}
We present a classical and quantum analysis of a particle confined in a three-dimensional paraboloidal cavity formed by two confocal paraboloids. Classically, the system is integrable and presents three independent constants of motion, namely, the energy, the $z$-component of the angular momentum, and a third dynamical constant associated with the paraboloidal geometry, which can be derived from the separability of the Hamilton--Jacobi equation. We derive closed-form analytical expressions for the actions, which allow us to determine the two conditions to get periodic closed trajectories. We classify these trajectories through the indices $(s,t,\ell)$. The caustic paraboloids that bound the motion provide a complete geometric characterization of admissible trajectories. Quantum mechanically, separability of the Schr\"odinger equation in parabolic coordinates yields eigenmodes described by Whittaker functions. We determine the energy spectrum and identify degeneracies arising not only from azimuthal symmetry but also from specific cavity deformations. A direct correspondence between classical trajectories and quantum eigenstates reveals that probability densities concentrate in the classically allowed region with controlled penetration into forbidden zones.
\end{abstract}

\maketitle

\date{\today}

\section{Introduction}

The characterization of particles confined within two-dimensional (2D) billiards and three-dimensional (3D) cavities has long provided a valuable framework for understanding fundamental aspects of classical and quantum dynamics \cite{TabachnikovBOOK,Berry1981,GutzwillerBOOK}. Depending on the geometry of the confining boundaries, these systems may exhibit integrable, mixed, or fully chaotic behavior \cite{GutzwillerBOOK,KozlovBOOK}. In the quantum domain, the same structures give rise to mode patterns, spectral statistics, and wave–classical correspondences that illuminate fundamental principles such as quantum ergodicity and scarring \cite{barnett2006,serbyn2021}. Establishing a clear connection between the classical and quantum descriptions remains central to several branches of contemporary physics, particularly for confined systems where geometry plays a central role.

Classical and quantum confocal parabolic billiards have been investigated for several years, revealing closed orbits, caustics, and separable eigenmodes~\cite{ParabolicBilliard,GravitationalParabolic,PhysRevE.109.034203}. They are integrable systems due to the separability of both the Hamilton-Jacobi and Schrödinger equations in parabolic coordinates \cite{Lopac,hillion1997, Fokicheva}. 
Quantum eigenstates are described by parabolic cylinder or confluent hypergeometric functions \cite{NISTBOOK,Hochstadt1986FunctionsMathPhysics}. Related parabolic cavities have also appeared in studies of atom optics, wave focusing, and electromagnetic resonators, and scattering ~\cite{PhysRevE.109.034203,Nockel}. These earlier works motivate a deeper exploration of fully three-dimensional confocal paraboloidal cavities, a system for which closed-form expressions and explicit classical–quantum correspondences have not been fully characterized until now.

In this paper, we formulate and analyze the classical and quantum dynamics of a particle confined by two confocal paraboloids that define a 3D paraboloidal cavity. We prove that the system is integrable and derive closed-form expressions for the dynamical actions in the Hamilton–Jacobi formalism, from which periodic trajectories follow by imposing rational winding conditions and azimuthal closure rules. In the process, we identify the three independent constants of motion of the particle and find its range of operation. We classify periodic orbits with three integer indices $(s,t,\ell)$ and describe their caustic structure in terms of constants of motion. For the quantum description, we solve the time-independent Schrödinger equation, obtaining separable eigenmodes labeled by three quantum numbers $(l,n,m)$ and demonstrating the existence of degeneracies arising not only from azimuthal symmetry but also from specific geometric ratios of the cavity boundaries. Finally, we relate quantum eigenstates to classical trajectories by pairing equal constants of motion. 

Beyond their intrinsic theoretical interest, paraboloidal cavities and related parabolic confinement geometries appear in several applied contexts. In nanotechnology, parabolic quantum dots and quantum corrals provide controllable platforms for studying quantum confinement and manipulating electron wave patterns at the nanoscale~\cite{Crommie1993,Manoharan2000,Crommie}. Parabolic micro-resonators and wave cavities are relevant to photonic devices, plasmonic structures, and atom surface interactions~\cite{ParabolicBilliard,Chang2006,Vahala2003}. Moreover, integrable geometries offer valuable benchmarks for semiclassical quantization and for testing numerical methods used in mesoscopic systems and nano-optics. The results presented here, particularly the closed-form expressions for actions, the classification of periodic trajectories, and the identification of spectral degeneracies, may thus be helpful for modeling and design of nanoscale resonators, quantum devices, and wave-confining structures with controlled geometric anisotropy.

\section{Classical mechanics formulation}

\begin{figure}[t] 
    \centering
    \includegraphics[width=13cm]{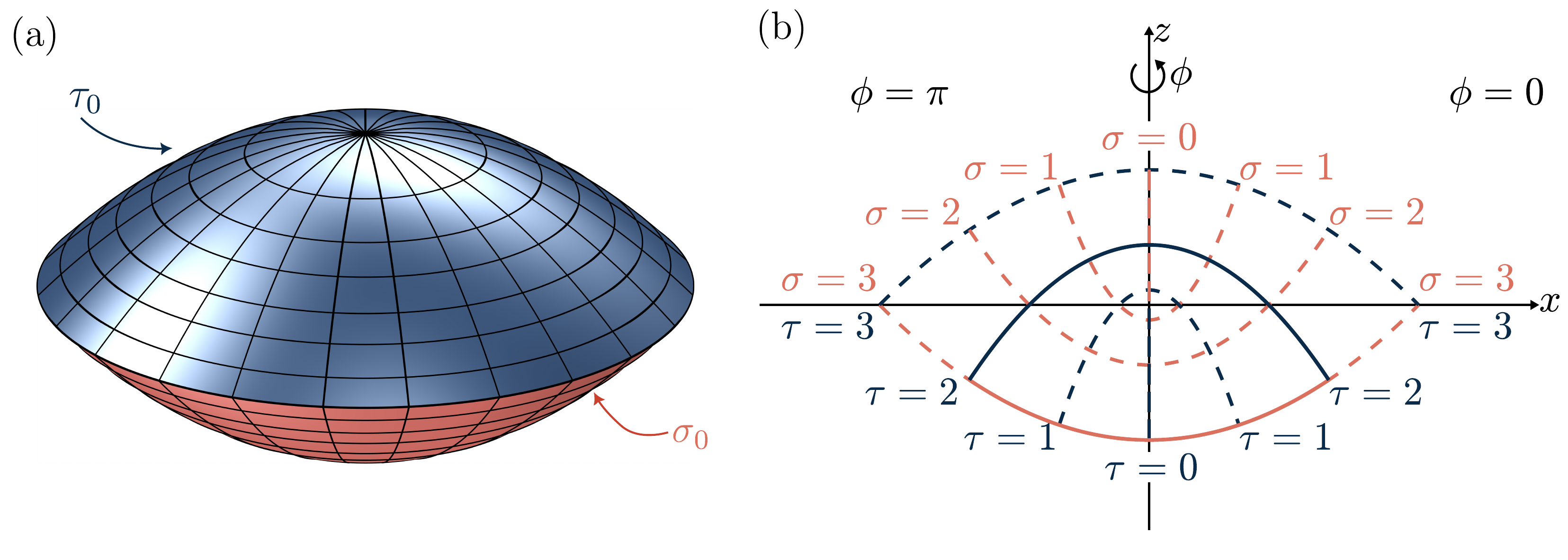}
    \caption{(a) Paraboloidal cavity defined by paraboloids $\sigma_0$ and $\tau_0$. (b) Parabolic coordinates $\mathbf{r} = (\sigma,\tau,\phi)$ defined in Eqs.~\eqref{pcs}. Surfaces of constant $\sigma$ and $\tau$ correspond to confocal paraboloids opening in the $z<0$ and $z>0$ directions, respectively.}
    \label{Figg1_Billiard}
\end{figure}
\subsection{Specifications of the cavity and coordinate systems}

Consider the motion of a point particle with mass $M$ confined within a confocal paraboloidal cavity, which is formed by two opposing paraboloids of revolution, as shown in Fig.~\ref{Figg1_Billiard}(a). The problem is naturally described in parabolic coordinates $\mathbf{r} = (\sigma,\tau,\phi)$ defined by
\begin{equation}
    x = \sigma\tau\cos\phi, \qquad \qquad
    y = \sigma\tau\sin\phi, \qquad \qquad
    z = (\tau^{2}-\sigma^{2})/2,
    \label{pcs}
\end{equation}
where $\sigma \in [0,\infty)$, $\tau \in [0,\infty)$, and $\phi \in [0,2\pi)$. The units of coordinates $\sigma$ and $\tau$ are the square root of length. Together with the azimuthal coordinate $\phi$, they form an orthogonal coordinate system with metric scale factors
\begin{equation}
    h_\sigma = h_\tau = \sqrt{\sigma^{2}+\tau^{2}}=\sqrt{2r}, 
    \qquad\qquad 
    h_\phi = \sigma\tau = \rho,
\end{equation}
where $r=(x^2+y^2+z^2)^{1/2}$ is the spherical radius and $\rho=(x^2+y^2)^{1/2}$ is the cylindrical radius.

Surfaces of constant $\sigma$ correspond to paraboloids opening in the $z<0$ direction, while surfaces of constant $\tau$ correspond to paraboloids opening in the $z>0$ direction, as shown in Fig.~\ref{Figg1_Billiard}(b). All paraboloids are confocal with focus at the origin.  

The boundaries of the cavity are specified by the paraboloids $\sigma=\sigma_{0}$ and $\tau=\tau_{0}$. The paraboloids intersect in a circumference of radius $\rho_0=\sigma_0 \tau_0$ placed at the plane $z_0 = (\tau_0^{2}-\sigma_0^{2})/2$. 
Note that if $\sigma_0$ and $\tau_0$ are multiplied by a common factor $c$, the cavity scales uniformly by a factor of $c^2$ without changing its shape.
Since the coordinates $(\sigma,\tau)$ are interchangeable, we can assume, without loss of generality, that $\sigma_0 \geq \tau_0$. 

The particle moves freely in straight lines and bounces elastically off the walls of the cavity. The momentum vector is written in Cartesian and parabolic coordinates as
\begin{equation}
\mathbf{p}=\left\{
\begin{array}
[c]{l}%
p_x~\hat{\mathbf{x}}+p_y~\hat{\mathbf{y}}+p_z~\hat{\mathbf{z}},\smallskip\\
\displaystyle
\frac{p_{\sigma}}{\sqrt{\sigma^{2}+\tau^{2}}}\hat{\boldsymbol\sigma}
+\frac{p_{\tau}}{\sqrt{\sigma^{2}+\tau^{2}}}\hat{\boldsymbol\tau}
+\frac{p_{\phi}}{\sigma\tau}\hat{\boldsymbol\phi},
\end{array}
\right.  
\qquad\qquad \left\vert \mathbf{p}\right\vert =P=\text{constant,}%
\end{equation}
where the canonical momenta $p_{\sigma}$ and $p_{\tau}$ have units of momentum by square root of length, and $p_{\phi}$ units of momentum by length, i.e., angular momentum. 

We assume that the particle bounces off the surfaces of the paraboloids, but not off the circumference formed by their intersection. This is because reflection is not defined at that intersection.

\subsection{Constants of motion}

Since the dynamical system is integrable, the particle possesses three independent constants of motion, namely:
\begin{enumerate}
\item The total energy $E=P^2/2M$, which is entirely kinetic. Conservation of energy implies conservation of the magnitude of the momentum; thus, we have
    \begin{equation}
        P^2 = 2ME = p_x^2+p_y^2+p_z^2
          = \frac{p_\sigma^2+p_\tau^2}{\sigma^2+\tau^2} 
          + \frac{p_\phi^2}{\sigma^2\tau^2}
          = \text{constant} > 0.
        \label{P2}
    \end{equation}
In the classical mechanics description, changing the kinetic energy of the particle affects only its speed, not its trajectory. Therefore, the shape of the trajectory within the cavity is determined entirely by the other two constants of motion.

\item The $z$-component of the angular momentum
    \begin{equation}
        L_z = x p_y - y p_x = p_\phi = \text{constant}.
        \label{Lz}
    \end{equation}
This condition comes from the rotational symmetry of the cavity about the $z$-axis. When $L_z$ is positive, the particle circulates counterclockwise around the $z$-axis when the trajectory is viewed from above. Conversely, if $L_z$ is negative, the circulation is clockwise. If $L_z = 0$, the trajectory is restricted to a meridional plane that always intersects the $z$-axis. In this situation, the problem reduces to the two-dimensional planar parabolic billiard \cite{ParabolicBilliard}. For later simplifications, we find it convenient to define a normalized constant of motion of the form
\begin{equation}
    \beta \equiv \frac{L_z^2}{P^2} = \frac{p_\phi^2}{P^2}= \text{constant} \geq 0,
    \label{beta}
\end{equation}
where $\beta$ has units of square length. 
\item As shown in Appendix A, the third constant of motion can be derived from the separability of the Hamilton–Jacobi equation. In Cartesian and parabolic coordinates, it is given by
\begin{equation}
C=\left\{
\begin{array}
[c]{l}%
2\left(p_x^2 + p_y^2\right)z - 2\left(xp_x + yp_y\right)p_z,\vspace{2mm} \\
\displaystyle \frac{p_{\sigma}^{2}-p_{\tau}^{2}}{2}+\left(\frac{\tau^{2}-\sigma^{2}}{2}\right)  \left( P^2 + \frac{p_{\phi}^{2}}{\sigma^{2}\tau^{2}}\right)
\end{array}
\right.
\label{AP}
\end{equation}
It is also convenient to define a normalized version of the constant $C$ of the form
\begin{equation}
     \alpha \equiv \frac{C}{P^2},
     \label{alfa}
\end{equation}
where $\alpha$ can be positive or negative, has units of length, and its range will be determined below.
\end{enumerate}

In what follows, the constants $P$, $\alpha$, and $\beta$ will be considered as the three independent constants of motion of the particle traveling inside the cavity.

\subsection{Canonical momenta, caustics, and trajectories}

By combining Eqs.~(\ref{P2}), (\ref{beta}), and (\ref{alfa}), we can decouple them and express the canonical momenta in terms of the conserved quantities $(\alpha,\beta,P)$ and the respective position variables as follows:
\begin{align} 
        p^2_\sigma &= P^2\,
        \left( \sigma^2 - \frac{\beta}{\sigma^2} + \alpha\right) \geq 0, \label{eq: p_sigma} \\
        p^2_\tau   &= P^2\,
        \left( \tau^2 - \frac{\beta}{\tau^2} -\alpha \right) \geq 0, \label{eq: p_tau} \\
        p^2_\phi &= P^2 \, \beta = \text{constant} \geq 0. \label{pphi}
\end{align}
As expected, the constant $P$ scales the three canonical momenta uniformly, preserving their proportions.

As the particle moves, the value of $p_\phi$ remains constant, but $p_\sigma$ and $p_\tau$ change continuously upon the coordinates $\sigma$ and $\tau$, always satisfying Eqs. (\ref{eq: p_sigma}) and (\ref{eq: p_tau}) for given $\alpha$, $\beta$, and $P$. 
Since $p^2_\sigma$ and $p^2_\tau$ cannot be negative, the triplets $(\sigma;\alpha,\beta)$ and $(\tau;\alpha,\beta)$ must satisfy the conditions
\begin{align}
    \sigma^4 + \alpha~\sigma^2 - \beta & \geq 0, 
    \qquad \text{for } \beta\geq0 \quad \text{and} \quad 0\leq\sigma\leq\sigma_0, \label{s4} \\
    \tau^4 -\alpha~\tau^2 - \beta  & \geq 0,
    \qquad \text{for } \beta\geq0 \quad \text{and} \quad 0\leq\tau\leq\tau_0, \label{t4}
\end{align}
simultaneously.

The condition for equality in Eq.~(\ref{s4}) occurs when $p_\sigma = 0$. At this point, the particle moves tangentially to a paraboloid described by $\sigma = \sigma_c$, which represents a caustic paraboloid. The motion of the particle is confined to the range $[\sigma_c, \sigma_0]$ within the cavity. Similarly, the condition for equality in Eq.~(\ref{t4}) results in the existence of a caustic paraboloid defined by $\tau = \tau_c$, which restricts the particle's motion to the range $[\tau_c, \tau_0]$ within the cavity. 
Setting the equality in Eqs.~(\ref{s4}) and (\ref{t4}) and solving for $\sigma^2$ and $\tau^2$, we get that the caustics are located at paraboloids
\begin{equation}
    \sigma_c = \sqrt{\frac{\Delta-\alpha}{2}},
    \qquad \qquad
    \tau_c   = \sqrt{\frac{\Delta+\alpha}{2}},
    \qquad\qquad
    \Delta \equiv \sqrt{\alpha^2+4\beta}.
    \label{caustics}
\end{equation}

Imposing the boundary conditions $0 \leq \sigma_c \leq \sigma_0$ and 
$0 \leq \tau_c \leq \tau_0$, we find that the admissible ranges of the constants $\alpha$ and $\beta$ are
\begin{equation}  \label{eq: Constants_Range}
-\sigma_0^2 \leq \alpha \leq \tau_0^2, \qquad \qquad
0 \leq \beta \leq \min(\sigma_0^4 + \alpha\sigma_0^2, ~\tau_0^4 - \alpha\tau_0^2). 
\end{equation}

\begin{figure}[t]
    \centering
    \includegraphics[width=16cm]{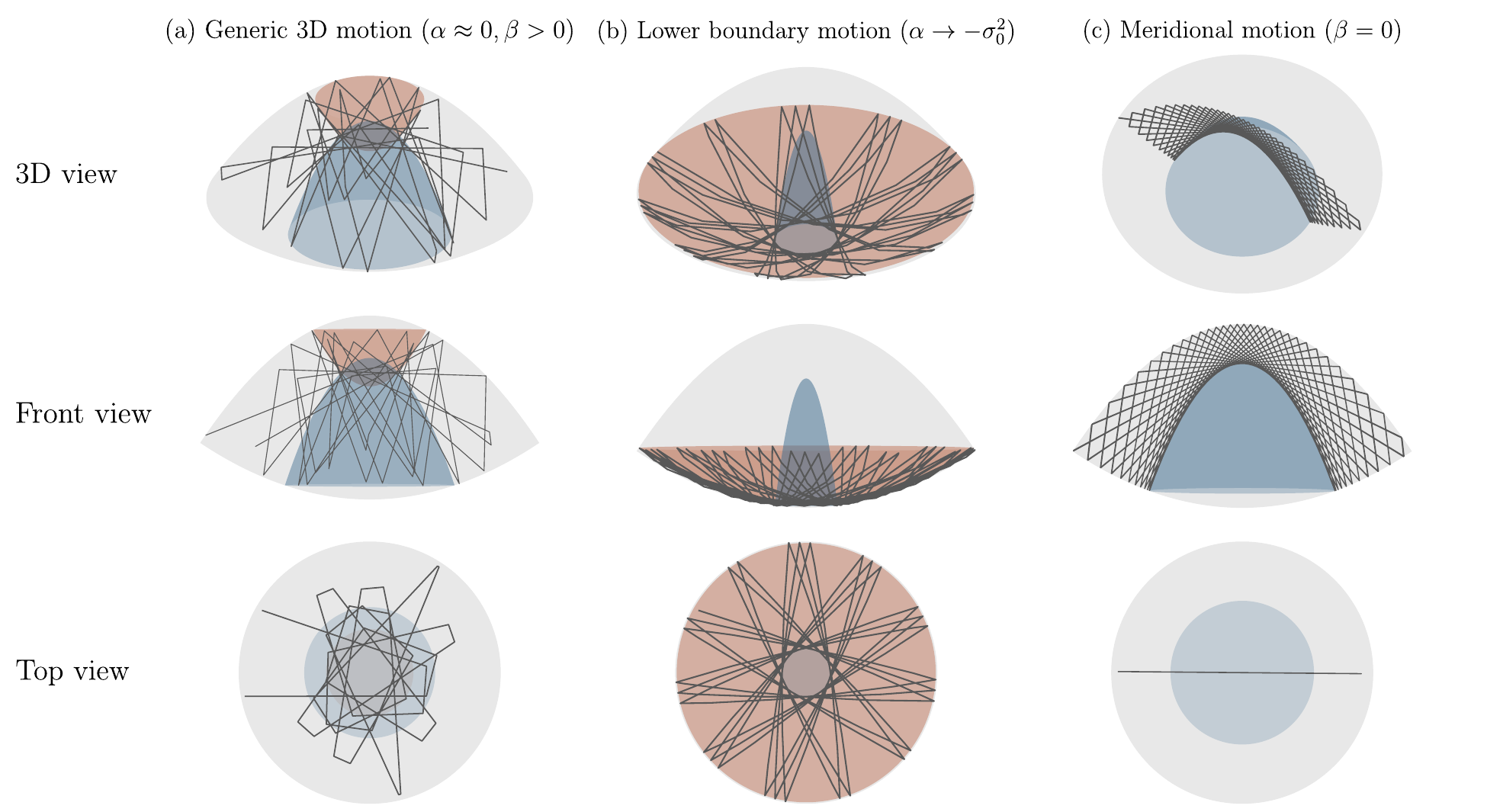}
    \caption{Classification of trajectories in the paraboloidal cavity. The blue-gray surface is the $\sigma$ caustic ($\sigma=\sigma_c$) and the apricot surface is the $\tau$ caustic ($\tau=\tau_c$). (a) Generic 3D motion with $\alpha\approx 0$ and $\beta>0$, bounded by both caustics. (b) Lower wall biased motion as $\alpha\to -\sigma_0^{2}$, with the $\sigma$–caustic approaching the boundary. (c) Meridional planar motion ($\beta=0$) that recovers the planar parabolic billiard. }
    \label{fig: Trajectories}
\end{figure}

Figure \ref{fig: Trajectories} illustrates three possible trajectories within the cavity. For each trajectory, a 3D view, a front view (from the $+x$-axis), and a top view (from the $+z$-axis) are provided to enhance visualization. The blue-gray and apricot surfaces represent the caustic paraboloids $\sigma_c$ and $\tau_c$, respectively. All segments of the trajectory are tangent to these paraboloids. 
Figure \ref{fig: Trajectories}(a) shows a generic trajectory with $\alpha$ close to zero and $\beta >0$. In this case, the upper and lower parts of the cavity are reached in approximately equal proportions. The particle moves in the region between the caustics and the boundary of the cavity. The particle can travel along the trajectory in either direction. In Eqs.~(\ref{eq: p_sigma})-(\ref{pphi}), reversibility is indicated by the sign of the momenta $(p_\sigma,p_\tau,p_\phi)$, which can be positive or negative.

For a given $\beta$, the parameter $\alpha$ mainly controls the vertical position of the orbit. As $\alpha \to -\sigma_0^2$, the motion stays close to the lower wall, as shown in Fig.~\ref{fig: Trajectories}(b); as $\alpha \to \tau_0^2$, it moves close to the upper wall. In both limits, the caustic becomes closer to the corresponding boundary, thus, the allowed region shrinks, and the particle practically moves tangentially to the wall.

A special case arises when angular momentum $L_z$ vanishes, i.e., $\beta=0$. In this case, the trajectory lies in a meridional plane, as shown in Fig.~\ref{fig: Trajectories}(c). 
For $\alpha>0$, the caustic opens downwards and the particle moves in the upper part of the cavity. The opposite situation occurs when $\alpha<0$. The case $\beta=0$ reduces to the planar two-dimensional parabolic billiard for which the conserved quantity reduces to
\begin{equation}
   C_{\beta=0} = \frac{\tau^2 p_\sigma^2-\sigma^2 p_\tau^2}{\sigma^2+\tau^2},
\end{equation}
which is proportional to the constant reported in Ref.~\cite{ParabolicBilliard}.

\subsection{Poincaré phase-space maps}

The dynamics of the particle within the cavity can be visualized on Poincaré maps by solving for the constants of motion $\alpha$ and $\beta$ from Eqs.~(\ref{eq: p_sigma}) and (\ref{eq: p_tau}). We get for the plane $(\sigma,p_\sigma)$
\begin{align}
\alpha(\sigma,p_\sigma;\beta,P) = \frac{p_\sigma^{\,2}}{P^2} - \sigma^{2} + \frac{\beta}{\sigma^{2}}, 
\hspace{15mm}
\beta(\sigma,p_\sigma;\alpha,P) = \sigma^{2}\!\left( \sigma^{2} - \frac{p_\sigma^{\,2}}{P^2} + \alpha \right), 
\label{alfabetaSS}
\end{align}
and for the plane $(\tau,p_\tau)$
\begin{align}
\alpha(\tau,p_\tau;\beta,P) = - \frac{p_\tau^{\,2}}{P^2} + \tau^{2} - \frac{\beta}{\tau^{2}},
\hspace{15mm}
\beta(\tau,p_\tau;\alpha,P) = \tau^{2}\!\left(\tau^{2} - \frac{p_\tau^{\,2}}{P^2} - \alpha \right). 
\label{alfabetaTT}
\end{align}

\begin{figure}[t]
    \centering
    \includegraphics[width=16cm]{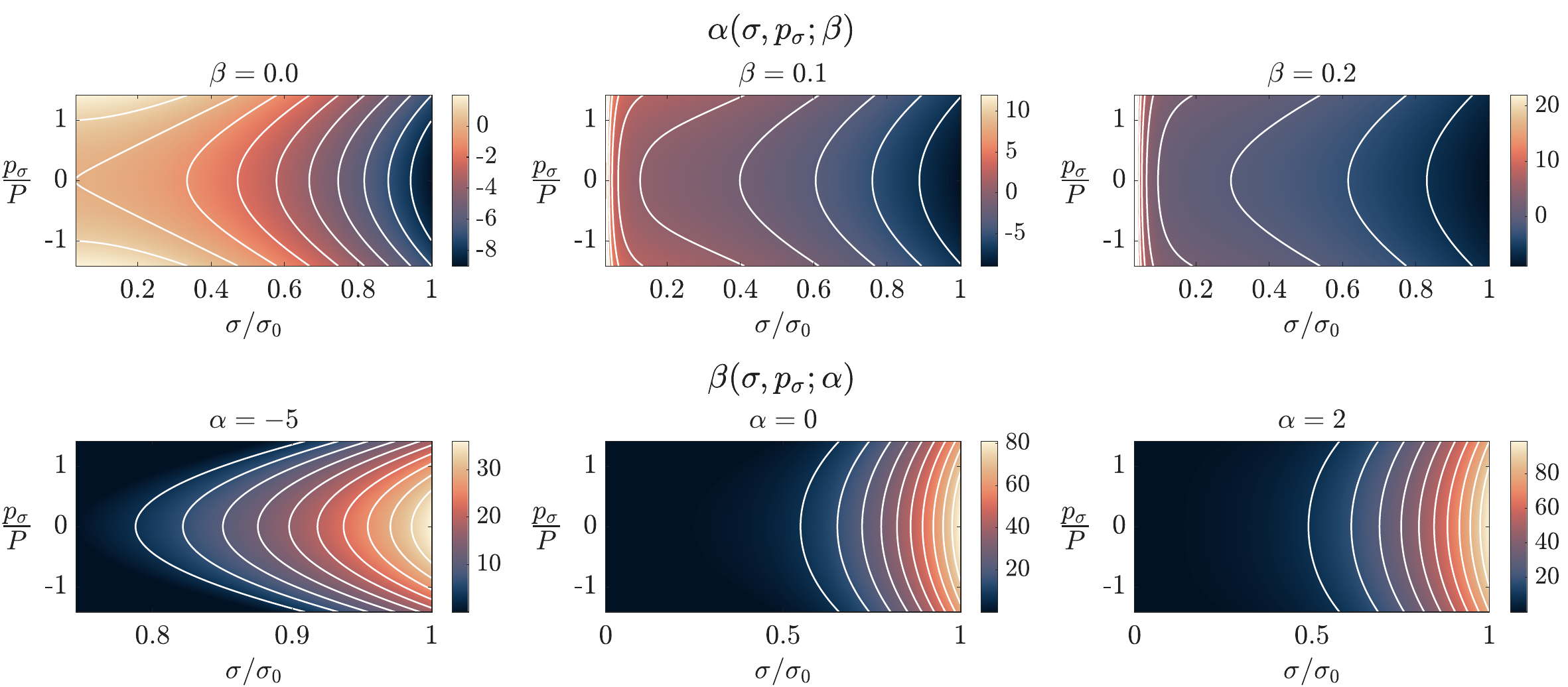}
    \caption{Poincaré phase maps $\alpha(\sigma,p_\sigma)$ and $\beta(\sigma,p_\sigma)$ for representative values $\alpha\in\{-5,0,2\}$ and $\beta\in\{0.0,0.1,0.2\}$ in a cavity with walls $\sigma_0=3$ and $\tau_0=2$.}
    \label{Figg3}
\end{figure}

In Fig.~\ref{Figg3}, we show the Poincaré maps $(\sigma,p_\sigma)$ of the paraboloidal cavity for several values of the constant $\alpha$ and $\beta$ by evaluating Eqs.~(\ref{alfabetaSS}). For given $(\beta,P)$, the curves  $\alpha(\sigma,p_\sigma;\beta,P)=\text{constant}$ are iso-$\alpha$ contours on the plane $(\sigma,p_\sigma)$. The remainder expressions are interpreted in the same way. 
All maps are even in momentum, $p_j\rightarrow -p_j$, so each contour curve is symmetric about $p_j=0$.
Similar plots can be generated for the plane $(\tau,p_\tau)$.

As the particle moves inside the cavity, the point defined by its phase-space coordinates $(\sigma,p_\sigma)$ moves along a contour curve where $\alpha$ and $\beta$ are constant, forming a closed loop that orbits clockwise. The motion of the point is restricted to the allowed regions defined by $\sigma \in [\sigma_c,\sigma_0]$, where the caustic is given by Eq.~(\ref{caustics}). When the point reaches the boundary $\sigma/\sigma_0=1$, it corresponds to an impact on the cavity wall, causing the momentum $p_\sigma$ to reverse. When the particle is tangential to the caustic, the point on the map crosses the $p_\sigma=0$ axis, reaching the minimum value of its coordinate $\sigma$.

The limit $\beta\to 0$ connects smoothly to the planar parabolic billiard \cite{ParabolicBilliard}. From Eq.~(\ref{alfabetaTT}) with $\beta=0$, we get
\begin{equation}
\alpha(\tau,p_\tau;0,P)=\tau^2-\frac{p_\tau^{\,2}}{P^2}.
\end{equation}
In this case the motion is entirely meridional, as illustrated in Fig. \ref{fig: Trajectories}(c). In contrast, the maps for $\beta(\sigma,p_\sigma;\alpha)$ given by Eq. (\ref{alfabetaSS}) demonstrate how the azimuthal barrier restricts access to lower values of $\sigma$. As $\beta\to 0$,  the admissible region becomes concentrated near $\sigma = 0$, which is consistent with the behavior observed in a planar billiard. In this case, trajectories can pass through $\sigma = 0$. However, for $\beta > 0$, the centrifugal term $\beta/\sigma^2$ excludes a neighborhood around $\sigma = 0$, which reflects the existence of two caustics in three-dimensional space.

For given $\beta$, changing $\alpha$ shears the $\alpha$-contours and moves the turning points $(\sigma_c,0)$ and $(\tau_c,0)$ along the $p_j=0$ axis. For given $\alpha$, increasing $\beta$ raises the centrifugal barrier, narrowing the allowed $\sigma$ interval and widening the allowed $\tau$ interval, consistent with the caustics in Fig.~\ref{fig: Trajectories}. Figure~\ref{Figg3} illustrates these trends: decreasing $\alpha$ shifts weight toward the lower wall by reducing $\sigma_c$ and increasing $\tau_c$, while increasing $\beta$ tightens the $\sigma$–contours near their turning segment and widens the $\tau$–contours. The 2D billiard map is recovered at the $\beta=0$ slice, and the progressive deformation with $\beta>0$ visualizes the transition from two-dimensional to three-dimensional motion.

\subsection{Actions and winding number function}

The action-angle variables in Hamilton-Jacobi theory \cite{FetterBOOK,GoldsteinBOOK} allow us to determine the conditions for having periodic trajectories within the cavity. The action associated with the canonical momentum $p_q$ is given by $J_q = (1/2\pi)\oint p_q~\d q$, where the integral is carried over a complete period of the coordinate $q$. 

The \textbf{actions} of the canonical momenta Eqs.~(\ref{eq: p_sigma})-(\ref{pphi}) are given by
\begin{align}
J_\sigma &= \frac{P}{\pi} \int_{\sigma_c}^{\sigma_0}\sqrt{\sigma^{2}-\frac{\beta}{\sigma^{2}}+\alpha}\, ~\d\sigma, 
   \label{Js}\\
J_\tau   &= \frac{P}{\pi} \int_{\tau_c}^{\tau_0}\sqrt{\tau^{2}-\frac{\beta}{\tau^{2}}-\alpha}\, ~\d \tau, 
   \label{Jt}\\
J_\phi   &= \frac{1}{2\pi}\int_0^{2\pi} p_\phi ~\d \phi = L_z = \text{constant}.
   \label{Jphi}
\end{align}

The integrals (\ref{Js}) and (\ref{Jt}) are essentially the same integral, but with the sign of $\alpha$ reversed. The action $J_\sigma$ can be evaluated in closed form; after some cumbersome simplifications, we get
\begin{equation}
J_\sigma(\sigma_0;\alpha,\beta)  
=\frac{P}{\pi}\left[
\frac{G}{2}
-\frac{\alpha}{4}\,
\ln\!\left(\frac{\alpha-2A}{\Delta}\right)
-\sqrt{\beta}\;
\arctan \left(
\frac{(\Delta+\alpha)\,A+2\beta}
{\sqrt{\beta}\,\bigl(\Delta+\alpha-2A\bigr)} \right)
\right],
\label{Jse}
\end{equation}
where
\begin{align}
    G &= G(\sigma_0;\alpha,\beta) \equiv \sqrt{\sigma_0^4+\alpha \sigma_0^2-\beta}, \\
    A &= A(\sigma_0;\alpha,\beta) \equiv G(\sigma_0;\alpha,\beta) - \sigma_0^2, \\
    \Delta & \equiv \sqrt{\alpha^2+4\beta}.
\end{align}

The expression for $J_\tau$ is equal to Eq.~(\ref{Jse}) with the substitutions $\sigma_0 \rightarrow \tau_0$ and $\alpha \rightarrow -\alpha$, that is
\begin{equation}
J_\tau = J_\sigma(\tau_0;-\alpha,\beta).
\label{Jte}
\end{equation}

Equations (\ref{Jse}) and (\ref{Jte}) provide the actions of the variables $\sigma$ and $\tau$, respectively. Although they are relatively long, it is remarkable that they admit a closed-form analytical expression in terms of elementary functions.

\vspace{3mm}

The \textbf{winding number function} $w(\alpha,\beta)$ is the ratio of the angle variables of the parabolic coordinates. For given $P$ and $\beta$, we have \cite{FetterBOOK,GoldsteinBOOK}
\begin{equation}
    w(\alpha,\beta) = \frac{\omega_\sigma}{\omega_\tau} = \frac{\partial H/\partial J_\sigma}{\partial H/\partial J_\tau} = \frac{\partial J_\tau}{\partial J_\sigma} = \frac{|\partial J_\tau / \partial \alpha|}{|\partial J_\sigma / \partial \alpha|},
\end{equation}
where $H$ is the Hamiltonian. 

The derivative of the actions $J_\sigma$ and $J_\tau$ can be calculated by directly differentiating (\ref{Jse}) and (\ref{Jte}), or by differentiating (\ref{Js}) and (\ref{Jt}) with respect to $\alpha$ and then integrating with respect to the variables. Both methods yield the same result:
\begin{equation}
    \frac{\partial J_\sigma}{\partial \alpha}
    =\frac{P}{4\pi} ~ \ln \left( \frac{2\sigma_0^2+\alpha+2G(\sigma_0;\alpha,\beta)}{\Delta} \right),
    \qquad\qquad
    \frac{\partial J_\tau}{\partial \alpha}
    =-\frac{P}{4\pi} ~ \ln \left( \frac{2\tau_0^2-\alpha+2G(\tau_0;-\alpha,\beta)}{\Delta} \right),    
\end{equation}
By replacing these results, we obtain
\begin{equation}
    w(\alpha,\beta) = \frac
    { \displaystyle \ln \left[ \frac{2\tau_0^2- \alpha +2G(\tau_0;-\alpha,\beta )} {\Delta} \right]}
    { \displaystyle \ln \left[ \frac{2\sigma_0^2+ \alpha +2G(\sigma_0;\alpha,\beta )}{\Delta} \right]}.
    \label{WN}
\end{equation}
Given the boundaries $(\sigma_0,\tau_0)$ of the cavity, the winding number depends only on the constants $\alpha$ and $\beta$. From Eqs.~(\ref{eq: Constants_Range}),  $w(\alpha,\beta)$ is valid within the triangle formed by the straight lines $\beta=0$,~ $\beta=\sigma_0^4+\alpha~\sigma_0^2$, and $\beta=\tau_0^4-\alpha~\tau_0^2$. The vertices of this triangle are located at the points $(-\sigma_0^2,0)$, $(\tau_0^2,0)$, and $(\tau_0^2-\sigma_0^2,\sigma_0^2 \tau_0^2)$ on the plane $(\alpha,\beta)$.

\subsection{Closed periodic trajectories}

For a particle to complete a closed periodic trajectory within the cavity, the following two conditions must be met simultaneously.

\begin{enumerate}

\item \textbf{Rational winding number.} The winding number Eq.~(\ref{WN}) is equal to a rational number, that is
\begin{equation}
     w(\alpha,\beta) = \frac{s}{t},\qquad s,t=\{1,2,3,\cdots\}.
     \label{wst}
\end{equation}
The trajectory closes after $s$ periods of the coordinate $\sigma$ and $t$ periods of the coordinate $\tau$. 

\item \textbf{Azimuthal periodic condition.} In one complete $(s,t)$ cycle, the net change in the azimuthal angle $\phi$ must be a multiple of $2\pi$. Contributions from single $\sigma$ and $\tau$ cycles follow from the action angle relations
\begin{equation}
    \Delta\phi_{\sigma} = \frac{\partial}{\partial L_z}\!\left( 2\pi J_\sigma \right) = 2\pi\,\frac{\partial J_\sigma}{\partial L_z},
    \qquad\qquad
    \Delta\phi_{\tau} =\frac{\partial}{\partial L_z}\!\left( 2\pi J_\tau \right) = 2\pi\,\frac{\partial J_\tau}{\partial L_z}.
\end{equation}
For the $(s,t)$ cycle, the orbit has $s$ oscillations in $\sigma$ and $t$ oscillations in $\tau$, resulting in a total azimuthal change
\begin{equation}
    \Delta\phi= 2\pi\!\left(s\,\frac{\partial J_\sigma}{\partial L_z}+ t\,\frac{\partial J_\tau}{\partial L_z} \right).
\end{equation}
A periodic trajectory requires that $\Delta\phi$ be equal to an integer multiple of $2\pi$, therefore 
\begin{equation}
    s\,\frac{\partial J_\sigma}{\partial L_z}+ t\,\frac{\partial J_\tau}{\partial L_z}= \ell, 
    \qquad \qquad \ell = \{ 0,1,2,\cdots ,\ell_{\max} \}, 
\end{equation}
where $\ell_{\max}$ will be determined below. Recalling from Eq.~(\ref{beta}) that $\beta = L_z^{2}/P^2$, the derivative with respect to the angular momentum writes as $\partial/\partial{L_z} = (2\sqrt{\beta}/P)~\partial/\partial\beta$.
Thus, the azimuthal periodic condition becomes
\begin{equation}
\frac{2\sqrt{\beta}}{P}\left(s\,\frac{\partial J_\sigma}{\partial \beta}+ t\,\frac{\partial J_\tau}{\partial \beta}\right)= \ell.
\label{acr1}
\end{equation}
The derivatives of the actions (\ref{Jse}) and (\ref{Jte}) with respect to $\beta$ are 
\begin{equation}
  \frac{\partial J_\sigma}{\partial \beta} = -\frac{P}{2\sqrt{\beta}} \vartheta_\sigma,  
  \qquad \qquad
  \frac{\partial J_\tau}{\partial \beta} = -\frac{P}{2\sqrt{\beta}} \vartheta_\tau,
\end{equation}
where 
\begin{align}
    \vartheta_\sigma=\vartheta_\sigma(\alpha,\beta) &\equiv \frac{1}{\pi} \arctan\left(\sqrt{\frac{(\Delta+\alpha)(\Delta-A_\sigma)}{(\alpha-\Delta)(\Delta+A_\sigma)}}\right),  \hspace{15mm}  A_\sigma\equiv 2\sigma_{0}^{2}+\alpha,
    \\
    \vartheta_\tau=\vartheta_\tau(\alpha,\beta) &\equiv \frac{1}{\pi} \arctan\left(\sqrt{\frac{(\Delta-\alpha)(\Delta-A_\tau)}{(-\alpha-\Delta)(\Delta+A_\tau)}}\right), \hspace{13mm}  A_\tau\equiv 2\tau_{0}^{2}-\alpha.
    \label{eq:theta_defs}
\end{align}
After replacing these expressions in Eq.~(\ref{acr1}), the azimuthal periodic condition can be written in compact form as
\begin{equation}
    s\,\vartheta_\sigma+t\,\vartheta_\tau=\ell, 
    \qquad\qquad s,t = \{1,2,3,\cdots\}, \qquad \ell = \{ 0,1,2,\cdots ,\ell_{\max} \}.
    \label{eq:azimuthal_closure}
\end{equation}
Dividing by $t$ and noting in Eq.~\eqref{wst} that $s/t$ equals the winding number, the azimuthal condition becomes
\begin{equation}
    f(\alpha,\beta) = w(\alpha,\beta)~\vartheta_\sigma + \vartheta_\tau = \frac{\ell}{t}=\text{rational number.}
\end{equation}
The total number of bounces in a periodic trajectory is given by $s+t$. Therefore, the index $\ell$, which represents the number of revolutions around the $z$-axis in a closed orbit, cannot be arbitrarily large; there would not be enough bounces to complete all the revolutions. The upper bound of $\ell$ is $\ell_{\max}= \left\lfloor (s+t)/2\right\rfloor$. 

\end{enumerate}

For a trajectory with indices $(s, t, \ell)$, Eqs.~(\ref{wst}) and (\ref{eq:azimuthal_closure}) form a system of two nonlinear algebraic equations for the variables $\alpha$ and $\beta$. These values can be determined numerically using a two-dimensional Newton-Raphson algorithm. Once we have calculated the values of $\alpha$ and $\beta$, we set a starting point of the trajectory on the surface of the cavity, ensuring that it lies within the allowed region $\sigma\in[\sigma_c,\sigma_0]$ and $\tau\in[\tau_c,\tau_0]$. Any point within this allowed region is valid, except for the points on the circumference where both paraboloids intersect. Once the starting point and constants $\alpha$ and $\beta$ are known, the components of the momentum vector of the first segment of the trajectory are calculated with Eqs.~(\ref{eq: p_sigma})-(\ref{pphi}). 

\begin{figure}[t]
    \centering
    \includegraphics[width=16cm]{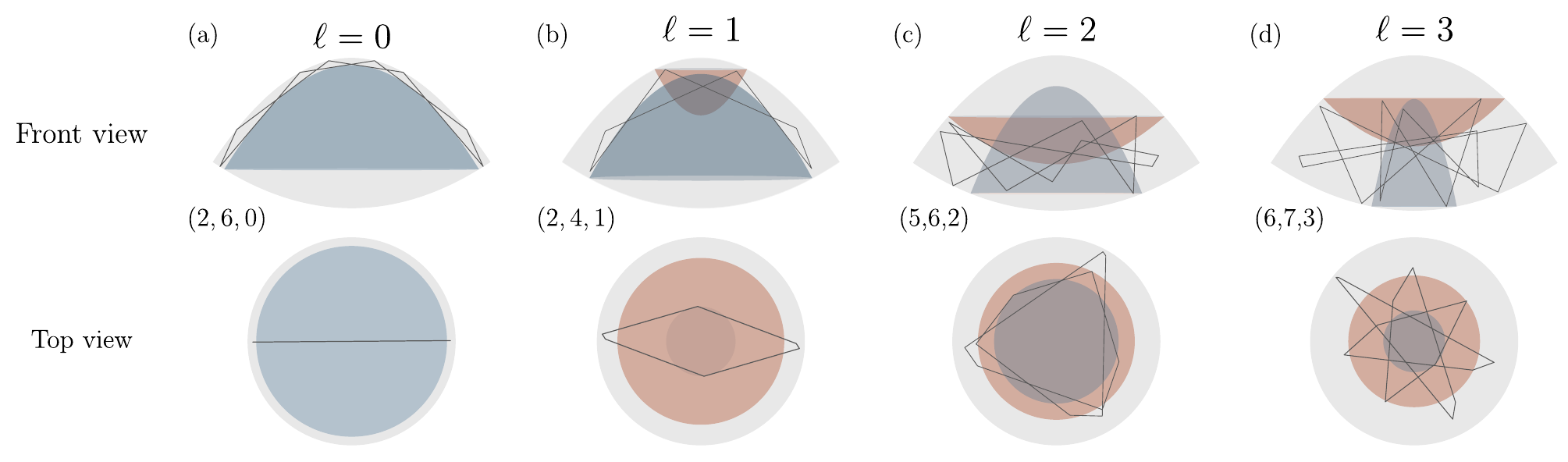}
  \caption{Periodic trajectories $(s,t,\ell)$ in the parabolic cavity with $\sigma_0=3$ and $\tau_0=2$. Panels (a)–(d) show representative examples with $\ell=0,1,2,3$, respectively. Here $s$ counts bounces on the $\sigma_0$ wall, $t$ counts bounces on the $\tau_0$ wall, and $\ell$ counts full revolutions about the $z$-axis.}
  \label{fig: Peridic_Trajectories}
\end{figure}

Figure~\ref{fig: Peridic_Trajectories} shows some periodic trajectories with $\ell = {0,1,2,3}$. For each trajectory, both a front view and a top view are provided for better visualization. The outer gray area represents the boundaries of the cavity, whereas the inner bluish and reddish regions indicate the paraboloidal caustics. The trajectories are always tangent to the caustics, and bounces occur off the walls.

During an orbit $(s,t)$, the particle reflects $s$ times at the paraboloid $\sigma=\sigma_0$ and $t$ times at the paraboloid $\tau=\tau_0$; thus, the total number of wall collisions per period is $s+t$. The index $\ell$ gives the number of revolutions the particle makes around the $z$ axis on the entire periodic trajectory. For the lowest case $\ell=0$, the angular momentum $L_z$ vanishes and the trajectory is confined to a meridional plane. Once $s$ and $t$ are defined, $\ell$ cannot be arbitrarily increased because the total number of bounces $(s+t)$ must be sufficient to ensure that the particle circulates $\ell$ times around the $z$-axis.
Given a specific trajectory with parameters $\alpha$ and $\beta$, the particle can orbit in either direction. In terms of motion constants, this is equivalent to reversing the sign of $L_z$.

The total length of a periodic trajectory $(s,t,\ell)$ with constants of motion $(\alpha,\beta,P)$ can be determined by applying Maupertuis’ principle \cite{FetterBOOK}: For fixed energy, the action satisfies $J=\int\sqrt{2M(E-U)}~\d l$, where $\d l$ is an element of path length. Therefore, we have
\begin{equation}
   L_{s,t,\ell}=\frac{2\pi}{P}(sJ_\sigma+tJ_\tau+\ell J_\phi),
\end{equation}
where the actions $J_\sigma$, $J_\tau$, and $J_\phi$ are given by Eqs.~(\ref{Jse}), (\ref{Jte}), and (\ref{Jphi}), respectively.

\section{Quantum mechanics formulation}

In the quantum regime, the stationary states $\psi(\sigma,\tau,\phi)$ of a particle of mass $M$ confined by the paraboloidal cavity are obtained from the time–independent Schrödinger equation
\begin{equation} \label{eq: Schrödinger}
\hat{H}\psi=\left[-\frac{\hbar^{2}}{2M}\nabla^{2}+U\right]\psi = E\,\psi,
\end{equation} 
with potential
\begin{equation}
U(\sigma,\tau,\phi)=
\begin{cases}
0,     \quad &(\sigma,\tau,\phi)\in\Omega,\\[2pt]
\infty,      &(\sigma,\tau,\phi)\notin\Omega,
\end{cases}
\end{equation}
where $\Omega=\{(\sigma,\tau,\phi):\,0\leq\sigma\leq\sigma_{0},\ 0\leq\tau\leq\tau_{0},\ 0\le\phi<2\pi\}$ is the interior of the confining walls $\sigma=\sigma_{0}$ and $\tau=\tau_{0}$. Eigenfunctions $\psi$ satisfy Dirichlet boundary conditions on the cavity surface $\psi(\sigma_{0},\tau,\phi)=0$ and $\psi(\sigma,\tau_{0},\phi)=0$.

In parabolic coordinates $(\sigma,\tau,\phi)$ the Laplacian reads
\begin{equation}
\nabla^{2}=\frac{1}{\sigma^{2}+\tau^{2}}\!\left(\frac{1}{\sigma}\frac{\partial}{\partial\sigma}+\frac{\partial^{2}}{\partial\sigma^{2}}+\frac{1}{\tau}\frac{\partial}{\partial\tau}+\frac{\partial^{2}}{\partial\tau^{2}}\right)+\frac{1}{\sigma^{2}\tau^{2}}\frac{\partial^{2}}{\partial\phi^{2}},
\end{equation}
and, inside the cavity, the equation reduces to the Helmholtz form
\begin{equation}
\nabla^{2}\psi+k^{2}\psi=0,\qquad\qquad k^{2}=2ME/\hbar^{2} = P^2/\hbar^{2}.
\label{hemholtz}
\end{equation}

\subsection{Stationary states of the paraboloidal cavity}

We seek separable solutions of the form $\psi(\sigma,\tau,\phi)=S(\sigma)\,T(\tau)\,\Phi(\phi)$. After replacing into the Schrödinger equation, the angular solution is given by
\begin{equation}
\Phi(\phi) = \exp(\ii m\phi),\qquad m\in\mathbb{Z},
\end{equation}
and the parabolic functions $S(\sigma)$ and $T(\tau)$ satisfy the pair of coupled equations
\begin{align}
    S''(\sigma)+\frac{1}{\sigma}S'(\sigma)+\left(k^{2}\sigma^{2}-\frac{m^{2}}{\sigma^{2}}+2ka\right)S(\sigma) &=0,\label{eq:S} \\
    T''(\tau)+\frac{1}{\tau}T'(\tau)+\left(k^{2}\tau^{2}-\frac{m^{2}}{\tau^{2}}-2ka\right)T(\tau) &=0, \label{eq:T}
\end{align}
where $a\in\mathbb{R}$ is the separation constant. 

Note that Eqs. (\ref{eq:S}) and (\ref{eq:T}) are the same equation, but with the sign of $a$ reversed. With an appropriate change of variables, these equations can be reformulated as the confluent hypergeometric equation. However, for our purposes, it is more convenient to express them as the Whittaker differential equation. To do this, we make the change of variables 
\begin{equation}
\zeta = \ii k\sigma^{2}, \qquad \chi = \ii k\tau^{2},
\end{equation}
and define the functions $U_S$ and $U_T$ as follows:
\begin{equation}
S(\sigma)=\sigma^{|m|}U_{S}(\sigma), \qquad T(\tau)=\tau^{|m|}U_{T}(\tau).
\end{equation}
With these substitutions, $U_{S}$ and $U_{T}$ satisfy the Whittaker differential equations \cite{NISTBOOK}
\begin{subequations}\label{eq:WhittakerPair}
\begin{align}
U_{S}''(\zeta) + \left(-\frac14 + \frac{\kappa_{S}}{\zeta} + \frac{\tfrac14 - \mu^{2}}{\zeta^{2}}\right)U_{S}(\zeta) &= 0, \hspace{10mm}  \kappa_{S} \equiv -\frac{\ii a}{2k},\label{eq:WhittakerPair:a}\\
U_{T}''(\chi) + \left(-\frac14 + \frac{\kappa_{T}}{\chi} + \frac{\tfrac14 - \mu^{2}}{\chi^{2}}\right)U_{T}(\chi) &= 0, \hspace{10mm}  \kappa_{T} \equiv +\frac{\ii a}{2k},\label{eq:WhittakerPair:b}
\end{align}
\end{subequations}
with $\mu=|m|/2$.

The solutions to Eqs. (\ref{eq:S}) and (\ref{eq:T}) read as
\begin{equation}
S(\sigma)=\frac{M_{\kappa_{S},\,\mu}(\ii k\sigma^{2})}{(\ii k)^{\mu}(\ii k\sigma^{2})^{1/2}}, \qquad \qquad
T(\tau)=\frac{M_{\kappa_{T},\,\mu}(\ii k\tau^{2})}{(\ii k)^{\mu}(\ii k\tau^{2})^{1/2}},
\label{ST}
\end{equation}
where $M_{\kappa,\mu}(z)$ is the Whittaker function \cite{NISTBOOK}. 

The cavity walls impose Dirichlet boundary conditions on $S$ and $T$, which discretize the allowed pairs $(k,a)$ and consequently the energy spectrum $E=\hbar^{2}k^{2}/(2M)$.
Collecting all partial results, the eigenstates are given by
\begin{equation}
    \psi_{l,n,m} = \mathcal{N}_{l,n,m} \, S_l(\sigma)\, T_n(\tau)\, \exp(\ii m\phi)
    \label{philnm}
\end{equation}
where $\mathcal{N}_{l,n,m}$ is a normalization constant such that $\int |\psi_{l,n,m}|^2 \, \d V = 1$, with the integration carried out over the whole volume of the cavity. 

The eigenstates $\psi(\sigma,\tau,\phi)$ must satisfy the Dirichlet boundary conditions $\psi=0$ at both walls. Evaluating Eqs.~(\ref{ST}) at $\sigma=\sigma_{0}$ and $\tau=\tau_{0}$ and setting them equal to zero gives the  conditions
\begin{align}
S(a,k;\sigma_0,m)&=0\quad\longrightarrow\quad M_{\kappa_{S},\,\mu}(\ii k\sigma_0^{2})=0, \label{Sso}
\\
T(a,k;\tau_0,m)&=0\quad\longrightarrow\quad M_{\kappa_{T},\,\mu}(\ii k\tau_0^{2})=0.\label{Tto}
\end{align}

\begin{figure}[t] 
    \centering
    \includegraphics[width=16cm]{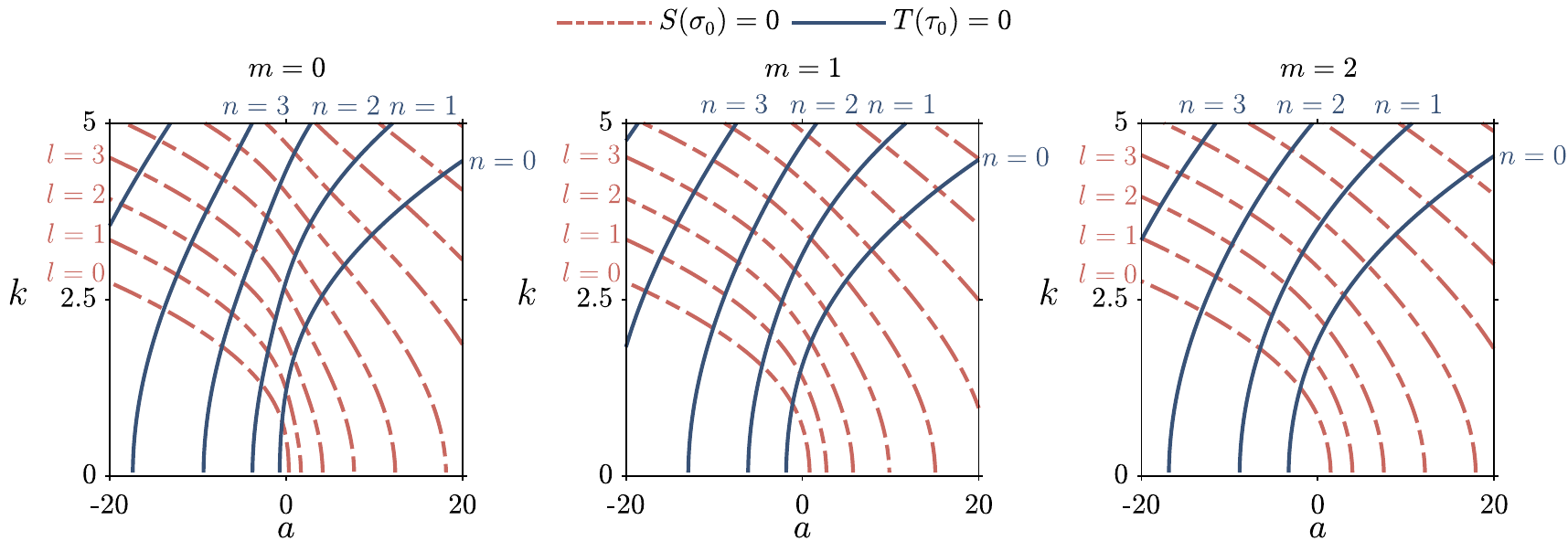}
    \caption{(a) Contour lines $S(\sigma_{0})=0$ and $T(\tau_{0})=0$ in the $(a,k)$ plane for $m= \{0,1,2\}$. Eigenpairs $(a,k)$ are given by the intersections of the two families.}
    \label{fig: NodalMap}
\end{figure}

Given the values of $(\sigma_0,\tau_0,m)$, conditions (\ref{Sso}) and (\ref{Tto}) form a system of two nonlinear equations for the variables $a$ and $k$. As shown in Fig.~\ref{fig: NodalMap}, each equation defines a set of zero contour lines in the plane $(a,k)$. The zero lines of the curves for $S(\sigma_0)$ and $T(\tau_0)$ are labeled with the indices $l$ and $n$, respectively. Each cross-point of these curves is associated with a specific combination of indices $(l,n,m)$, which in turn defines the parameters $(a_{l,n,m},k_{l,n,m})$ of the corresponding eigenstate $\psi_{l,n,m}$ with energy $E=\hbar^{2}k^{2}/2M$.

\begin{figure}[t] 
    \centering
    \includegraphics[width=14cm]{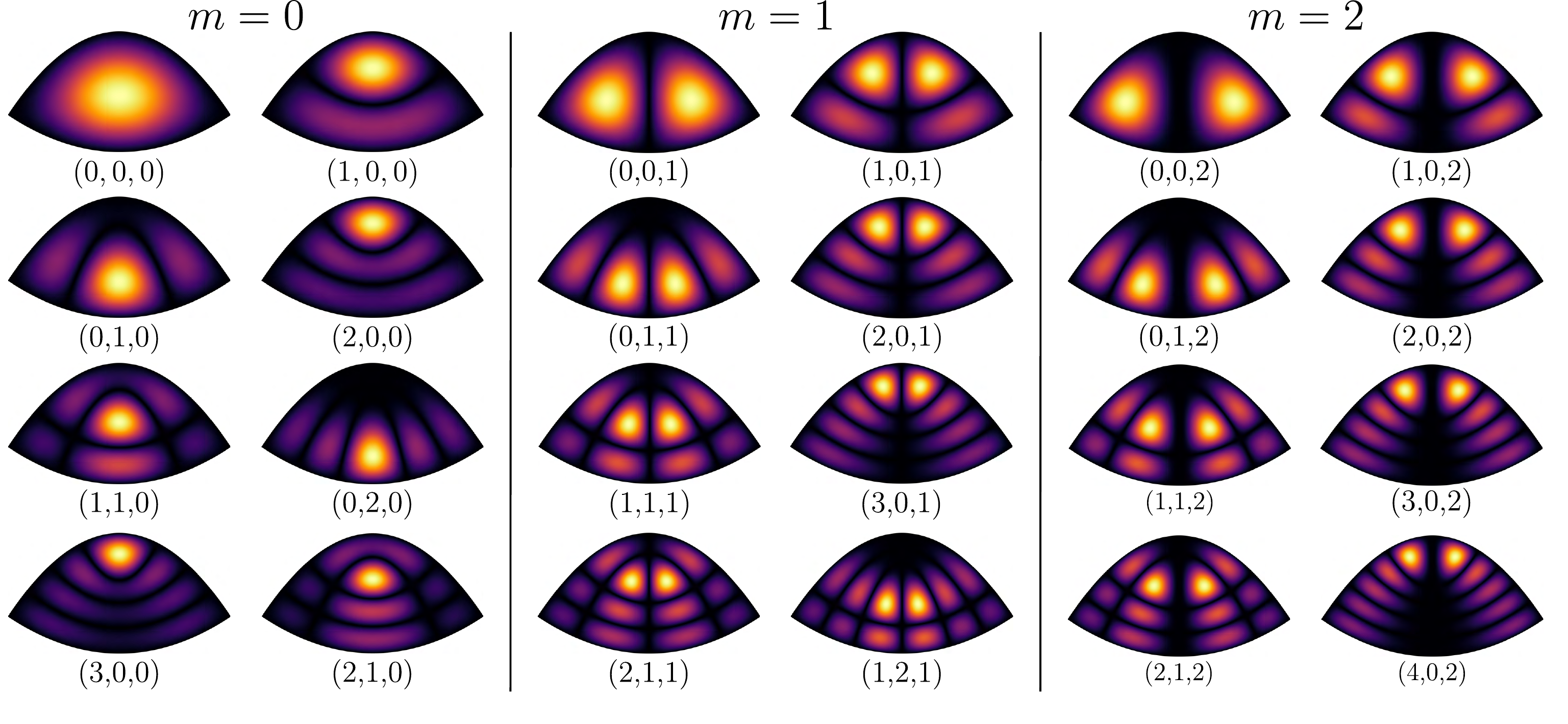}
    \caption{Eigenmodes $\psi_{l,n,m}$ for the families $m=\{0,1,2\}$, ordered by the energies $E_{l,n,m}$ for a paraboloidal cavity with $\sigma_{0}=3$ and $\tau_{0}=2$.}
    \label{fig: Modes}
\end{figure}

Figure \ref{fig: Modes} shows the probability distributions of the first eigenstates $\psi_{l,n,m}(\sigma,\tau,\phi)$ of the paraboloidal cavity with $\sigma_{0}=3$ and $\tau_{0}=2$. The patterns are plotted on a meridional plane, but have rotational symmetry about the $z$-axis.  We organize the eigenstates into three blocks with constant index $m = 0,1,2$. Within each block, the states are arranged in order of increasing energy. 
For $m = 0$, the states have no azimuthal dependence on $\phi$, and the distributions show non-zero values along the $z$-axis. However, for $|m| \geq 1$, the azimuthal dependence represented by $\exp(i m \phi)$ causes the eigenstate to be zero along the $z$-axis, as illustrated in Fig. \ref{fig: Modes}.

The eigenstates of the cavity form a complete orthonormal basis, that is
\begin{equation}
\iiint_{\Omega} \big[\psi_{l,n,m}(\mathbf{r})\big]^{*}\,\psi_{l',n',m'}(\mathbf{r})\, \d V
= \delta_{ll'}\,\delta_{nn'}\,\delta_{mm'} \, ,
\end{equation}
such that any wavefunction inside the cavity is a linear combination of states of the form Eq.~(\ref{philnm}).

\subsection{Energy spectrum and degeneracies}

The eigenstates in the paraboloidal cavity are degenerate because of the geometric properties of the cavity. First, the azimuthal symmetry makes the modes with $\pm m$ degenerate, i.e., $E_{l,n,+m}=E_{l,n,-m}$. When $m = 0$, the eigenstates are non-degenerate. On the other hand, if the boundaries are equal $\sigma_0=\tau_0$, the indices $(l,n)$ are interchangeable and the corresponding states are degenerate, that is, $E_{l,n,m}=E_{n,l,m}$.

\begin{figure}[t]
  \centering
  \includegraphics[width=10cm]{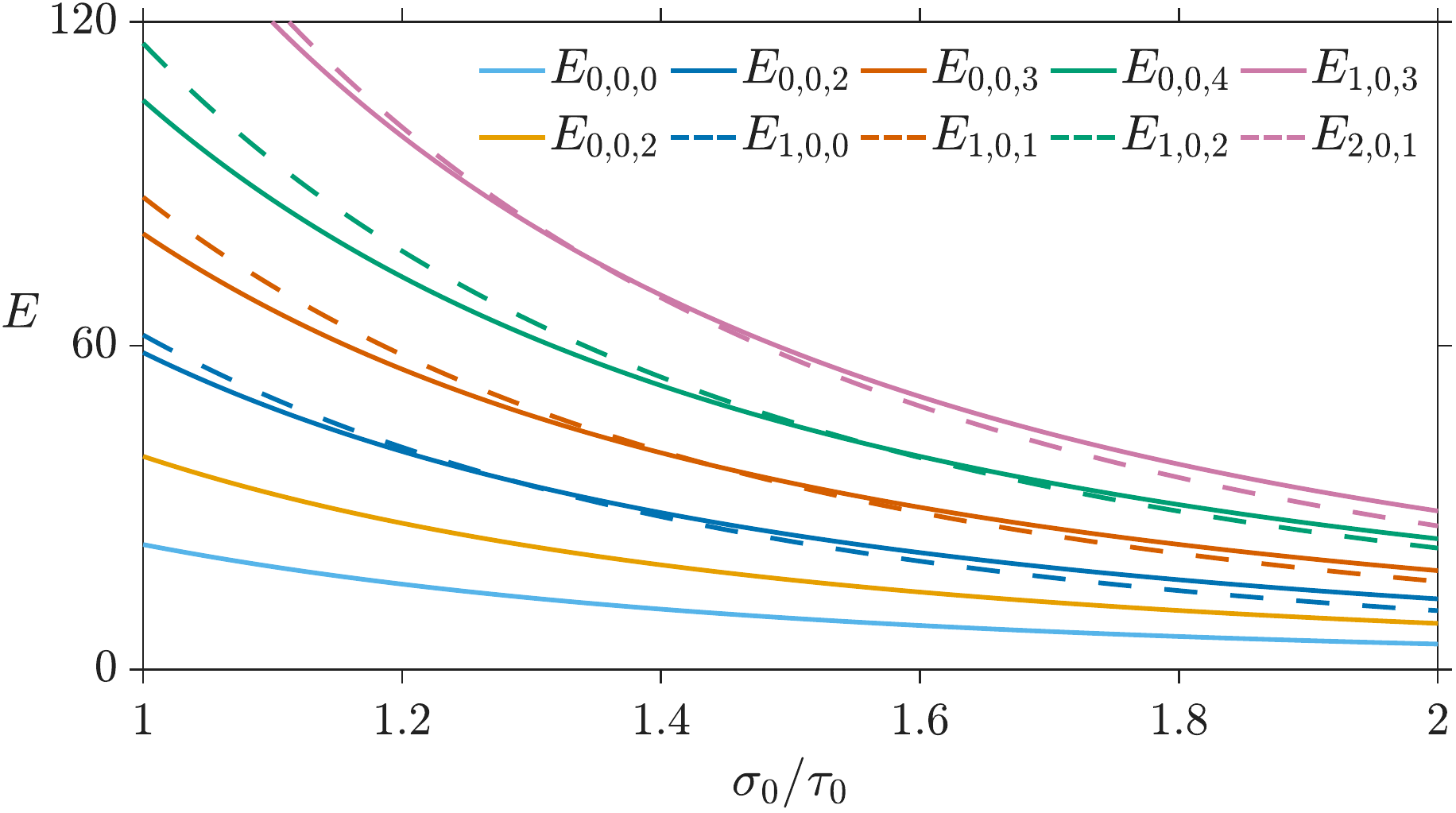}
  \caption{Energy spectrum $E_{l,n,m}$ (in units of $\hbar^{2}/2M$) as a function of the cavity shape $\sigma_{0}/\tau_{0}$ for fixed $\tau_{0}=1$. The plot shows that, as the cavity is deformed, energy levels associated with different index triplets $(l,n,m)$ intersect, leading to changes in the ordering by energy of the eigenstates.}

  \label{fig:eigensurface}
\end{figure}

More interestingly, the cavity can exhibit degenerate states even if the boundaries are not equal. To demonstrate this, in Fig. \ref{fig:eigensurface} we plot the energies of the first few eigenstates as a function of the cavity \textit{deformation} parameter, defined as $\sigma_0/\tau_0$. As can be seen, the lowest states (0,0,0) and (0,0,1) remain the least energetic throughout the interval. However, the curves of the states (0,0,2) and (1,0,0) intersect at $\sigma_0/\tau_0 \approx 1.25$. The same phenomenon occurs for other states. Thus, the order in which the eigenstates are ordered by energy also depends on the $\sigma_0/\tau_0$ ratio of the cavity.

\subsection{Constants of motion}

The same constants of motion found in classical mechanics should also be present in quantum mechanics. The separability of the Schrödinger equation in parabolic coordinates indicates the existence of three mutually commuting observables: the Hamiltonian, the $z$ component of angular momentum, and a second-order invariant related to the separation constant $a$. The energy eigenvalue equation is provided in Eq.~(\ref{eq: Schrödinger}). For the angular momentum, the operator correspondence
\begin{equation}
L_z \rightarrow - \ii \hbar \frac{\partial}{\partial \phi},
\end{equation}
yields the eigenvalue equation
\begin{equation}
\hat{L}_z \psi = -\ii \hbar \frac{\partial}{\partial\phi}\,\psi = m\hbar\,\psi.
\end{equation}
The third constant arises from the difference of the radial parts in Eqs.~(\ref{eq:S}) and (\ref{eq:T}). Using the operator substitutions
\begin{equation}
p_\sigma^2 \;\rightarrow\; -\hbar^2\!\left(\partial_\sigma^2 + \frac{1}{\sigma}\partial_\sigma\right),\qquad
p_{\tau}^2 \;\rightarrow\; -\hbar^2\!\left(\partial_{\tau}^2 + \frac{1}{\tau}\partial_{\tau}\right).
\end{equation}
one obtains the eigenvalue problem
\begin{equation}
   \hat{C}\,\psi = \frac{1}{2}
   \left[-\hbar^2\!\left( \partial_{\sigma}^2 + \frac{1}{\sigma}\partial_{\sigma}
- \partial_{\tau}^2 - \frac{1}{\tau}\partial_{\tau} \right)
+ \left(\frac{1}{\sigma^2}-\frac{1}{\tau^2}\right)L_z^2
+ 2M \hat{H}\,(\tau^2-\sigma^2)
   \right]\psi 
= 2\hbar^2ka ~\psi,
\end{equation}
where $\hat{H}$ is the Hamiltonian operator. Thus, for each eigenstate $\psi_{l,n,m}$,
\begin{equation}
L_{z\,;\,l,n,m} = m \hbar, \qquad \qquad C_{l,n,m} = 2\hbar^{2}k_{l,n,m}a_{l,n,m}.
\end{equation}

Together, the commuting set of operators $\{\hat{H},\hat{L}_z,\hat{C}\}$ resolves the spectrum into multiplets labeled by quantum numbers $(l,n,m)$, with $m\in\mathbb{Z}$ and $(l,n)\in\mathbb{N} $ determined by the nodal orders of $S$ and $T$ at the admissible intersections in the $(a,k)$ plane.

\subsection{Comparison between classical and quantum solutions}

A connection between classical trajectories and quantum eigenstates can be made through the constants of motion $\alpha$ and $\beta$. 
We can associate a quantum eigenstate to a classical trajectory if they have the same motion constants. To do this, we first select an eigenstate and calculate its constants $\alpha$ and $\beta$. Next, with these constants we determine the corresponding classical trajectory. If we are interested in associating a periodic trajectory with the eigenstate, then with $\alpha$ and $\beta$ we find the rational number $s/t$ that approximates the winding number to a given accuracy.
With $k$ defined by Eq.~(\ref{hemholtz}), the constants of motion for the quantum eigenstates are
\begin{equation}
    \alpha_{l,n,m} = \frac{C_{l,n,m}}{\hbar^2k_{l,n,m}^2} = \frac{2a_{l,n,m}}{k_{l,n,m}}, \qquad\qquad
    \beta_{l,n,m}  = \frac{L_{z\,;\,l,n,m}^2}{\hbar^2 k^2_{l,n,m}} = \frac{m^2}{k^2_{l,n,m}}.
\end{equation}

\begin{table}[t]
\centering
\caption{Eigenenergies (in units of $\hbar^2/2M$), constants $\alpha$ and $\beta$, and the penetration ratio $\Pi$  for a cavity with walls $\sigma_0=3$ and $\tau_0=2$.}
\label{Table: Eigenstates}
\begin{tabular}{ccccc}
\hline
$k^2$ & $\psi_{l,n,m}$ & $\alpha$ & $\beta$ & $\Pi$ \\
\hline
0.59 & $\psi_{0,0,0}$ & -1.17 & 0.00 & 0.06 \\
1.04 & $\psi_{0,0,1}$ & -2.18 & 0.96 & 0.13 \\
1.48 & $\psi_{1,0,0}$ & 0.04 & 0.00 & 0.01 \\
1.56 & $\psi_{0,0,2}$ & -3.15 & 2.56 & 0.20 \\
1.71 & $\psi_{0,2,0}$ & -4.28 & 0.00 & 0.25 \\
2.13 & $\psi_{1,0,1}$ & -0.37 & 0.47 & 0.10 \\
2.16 & $\psi_{0,0,3}$ & -4.09 & 4.16 & 0.31 \\
2.44 & $\psi_{0,2,1}$ & -5.85 & 0.41 & 0.56 \\
2.78 & $\psi_{2,0,0}$ & 1.02 & 0.00 & 0.27 \\
2.84 & $\psi_{0,0,4}$ & -5.01 & 5.64 & 0.49 \\
2.87 & $\psi_{1,0,2}$ & -0.91 & 1.39 & 0.13 \\
3.10 & $\psi_{1,2,0}$ & -2.34 & 0.00 & 0.26 \\
3.25 & $\psi_{0,1,2}$ & -7.24 & 1.23 & 0.88 \\
3.38 & $\psi_{0,1,0}$ & -7.90 & 0.00 & 0.96 \\
3.59 & $\psi_{0,0,5}$ & -5.93 & 6.97 & 0.73 \\
3.61 & $\psi_{2,0,1}$ & 1.03 & 0.28 & 0.22 \\
3.69 & $\psi_{1,0,3}$ & -1.51 & 2.44 & 0.17 \\
4.09 & $\psi_{1,2,1}$ & -3.52 & 0.24 & 0.45 \\
\hline
\end{tabular}
\end{table}


In Table \ref{Table: Eigenstates} we list the values of $(\alpha,\beta)$ for the eigenstates shown in Fig.~\ref{fig: Modes}, ordered by energy $k^2$.  As mentioned above, this ordering depends on the boundary parameters $(\sigma_{0},\tau_{0})$. Note that the range of the constants of motion in the quantum description is the same as in the classical description given by Eq.~(\ref{eq: Constants_Range}). Thus, an eigenstate is not associated with a single trajectory but rather with all trajectories that share its constants of motion. For this reason, to make the connection, it is easier to start from an eigenstate and compute its constants than to search for an eigenstate beginning from a particular classical trajectory.

\begin{figure}[b] 
    \centering
    \includegraphics[width=15cm]{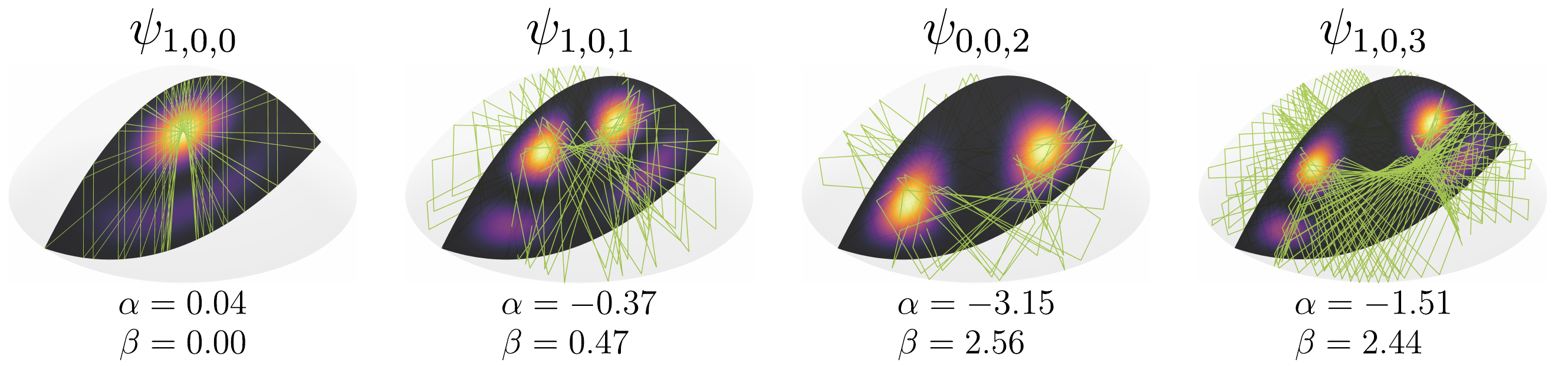}
   \caption{Wave amplitude $\psi_{l,n,m}$ and its associated classical trajectory, both characterized by the same constants of motion $\alpha$ and $\beta$.}
    \label{fig: Classical_Quantum}
\end{figure}

In Fig.~\ref{fig: Classical_Quantum}, we show the wave amplitudes of some eigenstates and their corresponding classical trajectories. 
We are unable to illustrate the full three-dimensional variation of the eigenstate, so instead, we represent its distribution on an inclined meridional plane. The classical trajectories are restricted to a region of space bounded by cavity walls and caustics.
Clearly, the probability distributions are more localized within the allowed region of the classical trajectory. Note that the quantum distribution
penetrates the forbidden region delimited by the classical caustics. To quantify this penetration, we calculate the ratio
\begin{equation}
    \Pi_{l,n,m} = \frac{\Pi_f}{\Pi_t} = \frac{\iiint_{\text{Forbidden volume}}|\psi_{l,n,m}(\sigma,\tau,\phi)|^2 \, \d V }{\iiint_{\text{Whole volume}} |\psi_{l,n,m}(\sigma,\tau,\phi)|^2 \, \d V },
\end{equation}
where $\Pi_f$ is the probability inside the classically forbidden region (defined by the inequalities implied by Eq.~(\ref{eq: Constants_Range}), $\Pi_t$ is the probability in the whole cavity, and $\d V=\sigma\tau(\sigma^2+\tau^2)\,\d\sigma\, \d\tau\, \d\phi$ is the volume element in paraboloidal coordinates. The values of $\Pi_{l,n,m}$ are listed in Table \ref{Table: Eigenstates}.

\section{Conclusions}

In this work, we formulated a complete classical and quantum description of a particle confined within a three-dimensional paraboloidal cavity bounded by two confocal paraboloids. On the classical side, we showed that the Hamilton-Jacobi equation separates in parabolic coordinates, leading to three constants of motion $(P,\alpha,\beta)$ that fully characterize the trajectories. A significant contribution is the derivation of closed-form analytical expressions for the actions $J_{\sigma}$ and $J_{\tau}$, written entirely in terms of elementary functions, which is remarkable given the nonlinear structure of the dynamical equations. These actions allow us to define the winding number $w(\alpha,\beta)$ and, together with the azimuthal closure condition, to obtain exact periodicity conditions for three-dimensional orbits in the cavity. The resulting families of periodic trajectories, parametrized by $(s,t,\ell)$, provide a complete classification of closed classical motions and naturally reduce to the known two-dimensional parabolic billiard when $\beta = 0$.

In addition to the analytical actions, the identification of the caustic paraboloids $\sigma_{c}$ and $\tau_{c}$ plays a key role in describing the particle motion. These caustics determine the admissible region of trajectories and the allowed ranges of $(\alpha,\beta)$. Altogether, the classical results confirm not only the integrability of the paraboloidal cavity but also explicit formulae for determining periodic trajectories.

On the quantum side, we showed that the stationary Schrödinger equation also separates in parabolic coordinates, yielding eigenmodes expressed in terms of Whittaker functions subject to Dirichlet boundary conditions. This separability leads to quantum numbers $(l,n,m)$ associated with the coordinates $\sigma$, $\tau$, and $\phi$, respectively. Besides the expected degeneracy produced by symmetries of the cavity, e.g., 
between modes with $\pm m$, and interchangeable states in symmetric cavities, an unexpected finding was the emergence of degenerate eigenstates for specific proportions $\sigma_0/\tau_0$ of the cavity. Thus, the spectral structure is affected by the geometric deformation of the cavity.

Finally, we established a direct classical-quantum correspondence by computing the constants of motion $(\alpha,\beta)$ for each quantum eigenstate and a set of classical trajectories. The comparison between caustics and probability densities confirms that quantum states localize predominantly within the classically allowed region while exhibiting finite penetration into the forbidden zones.



\bibliography{Parabolic3D}

\end{document}